\documentstyle[preprint,aps]{revtex}
\title{Antiproton production in Ni+Ni collisions at 1.85 GeV/nucleon}
\author{G. Q. Li and  C. M. Ko}
\address{Cyclotron Institute and Physics Department,\\
Texas A\&M University, College Station, Texas 77843}
\begin{document}
\maketitle

\begin{abstract}
Antiproton production in  Ni+Ni collisions at 1.85 GeV/nucleon
is studied in the relativistic Vlasov-Uehling-Uhlenbeck  model.
The self-energies of the antiproton are determined from
the nucleon self-energies by the G-parity transformation.
Also, the final-state interactions of the antiproton including
both rescattering and annihilation are explicitly treated.
With a soft nuclear equation of state,
the calculated antiproton momentum spectrum is in good
agreement with recent experimental data from the
heavy-ion synchrotron at GSI.
The effect due to the reduced nucleon and antinucleon masses
in a medium is found to be more appreciable than in earlier
Bevalac experiments with lighter systems and at higher energies.
\end{abstract}

The study of antiproton production in heavy-ion collisions at
subthreshold energies has been a topic of great interests
both experimentally
\cite{lbl,jinr,kek,gsi} and theoretically
\cite{dover,ko1,shor,dani,mos1,li1,li2,grei,mos2,li3}.
In Ref. \cite{li3}, we have studied antiproton production
from Si+Si collisions at 2.1 GeV/nucleon in the relativistic
Vlasov-Uehling-Uhlenbeck (RVUU) model \cite{ko2}.
Extending the RVUU model to include the antiproton degree of
freedom, we have been able to include the
medium effects on antiprotons and treat consistently their production,
propagation, rescattering, and
annihilation.  Because of the attractive scalar field,
both nucleon and antinucleon masses are reduced in a medium.
Assuming that the antiproton self-energies in a medium are given by the
G-parity transformation of the nucleon self-energies, then the
vector potential for the antiproton has an opposite sign from that
for the nucleon.  The vector potential
therefore does not play any role in antiproton production as the
antiproton
is produced together with a nucleon as a result of baryon conservation.
The reduction of nucleon and antinucleon masses in the medium
then reduces the antiproton production threshold and enhances thus the
primordial antiproton production in the dense matter formed
in nucleus-nucleus collisions. In Ref. \cite{li3}, our
theoretical results for the antiproton momentum spectrum
are found to agree with
experimental data from the Bevalac at Lawrence Berkeley Laboratory
\cite{lbl}.

Systematic measurements of antiproton production in heavier
systems and at lower incident energies are being carried out at the
heavy-ion synchrotron (SIS)
at GSI \cite{gsi}. Since the density reached in these collisions
is higher and the energy deficit is larger than in the
the experiments at Bevalac with lighter systems and at higher energies,
we expect that the medium effects
discussed above will be more appreciable in these collisions.
In this Brief Report, we
present our calculation of antiproton production in
Ni+Ni collisions at 1.85 GeV/nucleon and
compare the results with recent experimental data from  SIS \cite{gsi}.

The calculation is carried out in the same way as in Ref. \cite{li3}.
We use two sets of parameters for the nuclear equation of state
as given in Ref. \cite{li3}. They
lead to the same binding energy and nucleon effective mass at the
same saturation density but differ in the
incompressibility of the nuclear matter.
The one with an incompressibility of 200 MeV corresponds to
the soft equation of state, while the incompressibility of
the stiff equation of state is 380 MeV. We note that although the
nucleon effective mass ($m_N^*=~0.83 m_N$) at saturation density
is the same for the two set of parameters, it
differs at higher densities and is smaller
for the soft equation of state than for
the stiff equation of state (see Ref. \cite{li3}).

The nucleon effective mass also depends on the temperature of the medium.
In the Walecka model, the nucleon mass at finite density
is slightly increasing for the temperature range encountered in
heavy-ion collisions at SIS energies \cite{saito89}.
A similar temperature dependence of the constitutent quark mass
has also been found in calculations based on the Nambu$-$Jona-Lasinio
model \cite{kli90,bat94}.  This small change of the nucleon mass with
temperature is thus not too important.   It is, nevertheless, implicitly
included in our transport model as we determine at each time step and
for each cell the local effective mass as a function of local baryon
density and average kinetic energy. The latter is related to the
temperature in an equilibrium description.

Antiprotons are produced from $B_1B_2\rightarrow NNp\bar p$,
where $B_1$ and $B_2$ are either a nucleon or a delta. Higher baryon
resonances are not included in our calculation.  In Ref. \cite{grei},
subthreshold antiproton production has also been studied using the
relativistic quantum molecular dynamics where
all baryon resonances with masses below 2 GeV are included
However, no specific information is given regarding the importance
of the contribution from higher resonances relative to that from the delta.
It has been shown in Ref. \cite{alb} from the total photonuclear cross
section that the widths of higher baryon resonances increase
substantially already at normal nuclear matter density.
To treat these
broad and mostly overlapping baryon resonances
as elementary particles in a dense matter may thus be questionable.

Antiprotons can also be produced from pion-baryon interactions. In our
model pions are produced from the decay of deltas and thus materialize at the
later stage of heavy-ion collisions when the system already starts to
expand and baryons have therefore less kinetic energies.
Also, at the expansional
stage the effect of reduced nucleon mass is less significant
because the density is not high. The chance of
a pion-baryon collision with sufficient energy to produce a proton-antiproton
pair is thus smaller than that from the energetic baryon-baryon collisions
occurring at the compressional stage of heavy-ion collisions.
In Ref. \cite{grei}, the situation is different as
mesons (including higher-mass ones in addition
to pions) are produced from string breaking and thus are
present already at the early stage of heavy-ion collisions.
The meson-baryon contribution in Ref. \cite{grei} is about 50\%
of the total antiproton yield.   We expect that the pion-baryon
contribution in our approach is about 20-30\% of the total antiproton
yield, similar to the pion-baryon contribution to subthreshold
kaon production \cite{xio90,li94}.
As the medium effects discussed in this paper
are more significant than the pion-baryon contribution, we shall neglect
the latter in present work.

In the free space, the antiproton production cross section
from the process $B_1B_2\rightarrow NNp{\bar p}$
has been parametrized in Refs. \cite{mos1,li3} by
$\sigma _{B_1B_2}^{\bar p}(\sqrt {s})=0.012~(\sqrt {s}-\sqrt{s_0})^{1.846}$,
where $\sqrt{s}$ is the center-of-mass energy of the colliding baryons
and $\sqrt{s_0}=4m$ is the antiproton production threshold energy.
In a medium, the reduced nucleon and antinucleon masses should enter in
the antiproton production cross section not only into $\sqrt{s}$ and
$\sqrt{s_0}$ but also into other parameters.
Due to the lack of knowledge of the
process in a nuclear medium and
for an exploratory study of antiproton production in heavy-ion collisions,
we assume that the cross section has the same
form as that in free space, with
corresponding energy and threshold replaced by the medium-dependent ones, i.e.,
\begin{eqnarray}
\sigma _{B_1B_2}^{\bar p}(\sqrt {s^*})=0.012~(\sqrt {s^*}-\sqrt
{s^*_0})^{1.846}.
\end{eqnarray}

We  note that, within the mean-field approximation
and under the G-parity transformation,
the vector potential energy is the same in both the initial and the final
state of the reaction $B_1B_2\rightarrow NNp{\bar p}$ and thus
does not play any role in antiproton production.
The total center-of-mass energy $\sqrt {s^*}$ of the colliding pair of
baryons is thus given by
\begin{eqnarray}
\sqrt {s^*}=({\bf p}_1^*+m_1^*)^{1/2}+({\bf p}^*_2+m^*_2)^{1/2},
\end{eqnarray}
where $m_i^*$ and ${\bf p}^*_i$ ($i$=1,2) are, respectively,
the effective mass and kinetic momentum of the colliding baryons.
The antiproton production threshold in the medium is
\begin{eqnarray}
\sqrt {s^*_0}=4m^*.
\end{eqnarray}
In the reaction $B_1B_2\rightarrow NNp\bar p$, about
twice nucleon effective mass [Eq. (2)] is involved in the initial state
while four times nucleon effective mass [Eq. (3)] appears in the
final state, the
decrease of the nucleon and antinucleon masses in dense medium
leads to a reduction of the threshold and thus
an enhanced production of antiprotons from heavy-ion collisions.

In writing down Eqs. (2) and (3) we include only mean-field
contribution to the antiproton self energies. In principle, the dispersive
correction due to both antiproton elastic scattering and annihilation
by a nucleon should be added to the mean-field contribution.
In Ref. \cite{mos2}, the dispersive correction from antiproton annihilation
has been evaluated and is found to be appreciable.
However, the dispersive correction
involves higher-order loop diagrams which is expected to be somewhat
suppressed in heavy-ion collisions due to the highly nonequilibrium
nature of the dynamics.  As
an exploratory study and since no dispersive corrections have been added
to either the nucleon or the pion mean-field potential in the
transport model,
we neglect thus the dispersive contribution to antiproton potential
in the present calculation and plan to address this question in the future.

The final-state interactions of primordial antiprotons with
baryons are explicitly treated in our calculation. These include the
propagation of antiprotons in the mean-field potential and their
elastic rescattering and annihilation by baryons.
The mean-field potential is determined from the self-energies of the
antiproton. For both elastic scattering and annihilation, the
cross sections in free space as parametrized in Ref. \cite{cug1}
are used in the calculation.

The theoretical results shown below are obtained with the soft equation
of state and including all medium effects, unless otherwise explicitly
stated. In Fig. 1 we show the antiproton abundance as a function of time
for a head-on (b=0 fm) Ni+Ni collision at 1.85 GeV/nucleon.
The dashed and the solid curve give the primordial and the
final antiproton (after taking into account
annihilation) abundance, respectively.
The primordial antiproton abundance is about 1.7$\times 10^{-5}$ but
is reduced to about 2.2$\times 10^{-7}$ due to annihilation. So
only about 1.3\% of primordial antiprotons can escape from the dense hadronic
matter and be detected. Comparing with the results from the Si+Si collision
\cite{li3}, we find that the annihilation effect is more appreciable in
a heavier system like Ni+Ni. We also show in this figure
by the dotted curve the time evolution
of the central density $\rho /\rho _0$ ($\rho _0=~0.17~fm^{-3}$).
It is clearly seen that antiprotons are produced in the high
density region where the reduction of the production threshold is most
appreciable. As in the Si+Si collision \cite{li3},
antiprotons are mainly produced from the nucleon-delta interaction.

The comparison of our theoretical results with the experimental data is given
in Fig. 2, where the antiproton production cross section
is plotted as a function of the antiproton momentum in the laboratory.
The dashed
curve gives the results for the primordial antiprotons, while the solid
curve is the final antiproton spectrum with all final-state interactions
(propagation in the mean-field potential, elastic rescattering, and
annihilation) taken into account. The recent experimental data from
SIS \cite{gsi} are shown in the figure by solid circles.
It is seen that
the theoretical results are in reasonable agreement with
the data \cite{gsi}. The calculated cross section at $p_{\rm lab}=~1.0$
GeV/c is somewhat below the experimental value which has, however,
a large error.

The effect of the nuclear equation of state on antiproton
production is shown in Fig. 3.
It is seen that the antiproton
production  cross section  with the stiff equation of
state is about  a factor of 2-3 smaller than that with the soft equation of
state and is thus below the experimental data by
the same factor. This
is due to the larger incompressibility
(thus a larger compressional energy) and effective mass
at high densities (thus a higher threshold) in a
stiff equation of state than those in a soft equation of state.
The effect due to the equation of state is more significant than in
the Si+Si collision at 2.1 GeV/nucleon where we have found that
the  antiproton yield using a soft equation of state is
only about 50\% larger
than that using a stiff equation of state \cite{li3}.
Given the uncertainties in the
antiproton production and annihilation
cross sections in a medium, it is, however,
premature to conclude that the soft
equation of state is favored by the antiproton data from
heavy-ion collisions at subthreshold energies.

The reduction of in-medium nucleon and antiproton masses is expected to
lead to an enhancement of primordial antiproton production
as a result of the decreasing production threshold.
To see this  explicitly,
we have carried out three calculations for the antiproton production
cross section.
The first calculation is done
in the usual non-relativistic VUU model, using a soft Skyrme
parametrization for the equation of state ($K=200$ MeV) \cite{aich}.
In this case, baryon masses do not change with density as
the mean-field potential is momentum independent. Bare nucleon
and antiproton masses are thus used in Eqs. (1)-(3). The result
is shown in Fig. 4 by the dotted curve. The other two calculations
are carried out in the  RVUU model. In one calculation, the
bare antiproton mass is used, i.e., only the nucleon mass decreases with
density. The threshold in this case is thus 3$m^*+m$. The result
of this calculation is shown in Fig. 4 by the dashed curve. The
antiproton production cross section in this case (dashed curve)
is enhanced by about a factor of 12 over the
result with the  bare nucleon mass (dotted curve).
In the final calculation, both
the nucleon and the antiproton effective mass are used and the
threshold is
therefore 4$m^*$. The result is shown in Fig. 4 by the solid curve.
It is seen that the antiproton production cross section is further
enhanced by about a factor of 8 as compared to the second case.
Overall, the antiproton production
cross section is enhanced by about two orders of magnitude
due to the reduction of baryon masses
in a medium. In Ref. \cite{li3}, we have found that for Si+Si collisions
at 2.1 GeV/nucleon, the overall enhancement factor of antiproton yield
due to dropping nucleon and antinucleon masses in a medium is about 20.
The medium effects are thus
more clearly seen in heavy-ion collisions with heavier
systems and at lower incident energies.

Antiproton production in
the Ni+Ni collision at 1.85 GeV/nucleon has also been
studied by Teis {\it et al}. \cite{mos2} based on a relativistic transport
model that is similar to ours. They need, however, only a moderate attractive
antiproton potential to account for the experimental data.  This is
due to the smaller nucleon effective mass in their calculation than in ours
as a result of a stronger attractive scalar potential
at high densities.  Since the properties of a nucleon at
high densities have not been well determined, whether
the antiproton has a strong
attractive potential in dense medium is still an open question, and
more theoretical study is therefore needed.

In summary, we have calculated the
antiproton production cross section in the Ni+Ni
collision at 1.85 GeV/nucleon within the relativistic Vlasov-Uehling-Uhlenbeck
model which has been extended to include the antiproton degree of freedom.
The nucleon self-energies have been calculated in the non-linear sigma-omega
model, while the antiproton self-energies are obtained from the nucleon
self-energies by the G-parity transformation. Because of the
attractive scalar potential, both
nucleon and antiproton masses decrease with increasing density.
The antiproton
final-state interactions with baryons have been explicitly treated in
the calculation. With a soft equation of state,
our theoretical results are in good agreement
with recent experimental data from the SIS at GSI \cite{gsi}.
Our study confirms thus the conclusion of Ref. \cite{li3}
that it is essential to include
the attractive scalar potentials for both nucleon and antinucleon
in accounting for the measured antiproton yield.

\vskip 1cm

This work was supported in part by the National Science Foundation under
Grant No. PHY-9212209 and the Welch Foundation under Grant No. A-1110.

\pagebreak

\begin{figure}
\caption{Time evolution of antiproton abundance and central density.
The dashed and the solid
curve correspond to the primordial and the final (with antiproton
annihilation effect included) antiproton abundance,
respectively. The dotted curve gives the central density.}
\end{figure}

\begin{figure}
\caption{Antiproton momentum spectrum at $\theta _{lab}=0^0$ in
a Ni+Ni collision at 1.85 GeV/nucleon. The dashed curve is for the primordial
antiprotons, while the solid curve is the final result obtained
with antiproton
propagation, elastic rescattering, and annihilation.
The experimental data are taken from Ref. [4].}
\end{figure}

\begin{figure}
\caption{Same as Fig. 2. The solid and the dashed curve are the
results obtained with the soft and the stiff equation of state, respectively.}
\end{figure}

\begin{figure}
\caption{Same as Fig. 2. The
dotted curve gives the results obtained from the non-relativistic
VUU calculation. The results obtained from the RVUU
calculation with the bare antiproton mass are given by the dashed curve.
The solid curve gives the
results of the RVUU calculation with both the nucleon and
the antiproton in-medium mass.}
\end{figure}

\end{document}